# Triple-cation perovskite solar cells fabricated by hybrid PVD/blade coating process using green solvents

*Severin Siegrist, Shih-Chi Yang, Evgeniia Gilshtein, Xiaoxiao Sun, Ayodhya N. Tiwari and Fan Fu\**

Laboratory for Thin Films and Photovoltaics, Empa – Swiss Federal Laboratories for Materials Science and Technology, Ueberlandstrasse 129, 8600 Duebendorf, Switzerland

*Email: fan.fu@empa.ch

ABSTRACT

The scalability of highly efficient organic-inorganic perovskite solar cells (PSCs) is one of the remaining challenges of solar module manufacturing. Various scalable methods have been explored to strive for uniform perovskite films of high crystal quality on large-area substrates. However, each of these methods have individual drawbacks, limiting the successful commercialization of perovskite photovoltaics. Here, we report a fully scalable hybrid process, which combines vapor- and solution-based techniques to deposit high quality uniform perovskite films on large-area substrates. This two-step process does not use toxic solvents, and it further allows facile implementation of passivation strategies and additives. We fabricated PSCs based on this process and used blade coating to deposit both charge transporting layers ($SnO_2$ and Spiro-OMeTAD) without hazardous solvents in ambient air. The fabricated PSCs have yielded open-circuit voltage up to 1.16 V and power conversion efficiency of 18.7 % with good uniformity on 5 cm x 5 cm substrates.

**Introduction**

Organic–inorganic halide perovskite solar cells (PSC) have shown tremendous advancements in the past ten years, now reaching certified power conversion efficiency (PCE) of 25.5 % [1]. This rapid development can be mainly attributed to the excellent optoelectronic properties of the perovskite material and to the development of facile perovskite fabrication [2, 3, 4, 5, 6, 7]. One of the remaining challenges towards successful commercialization of perovskite photovoltaics is to fabricate every layer of the PSC by scalable deposition methods and to demonstrate high performance devices on large-area. Generally, these scalable deposition methods can be divided into solution- and vapor-based techniques. Solution-based methods, such as blade coating [8, 9], slot-die coating [10, 11], inkjet printing [1, 12], spray coating [13] offer the possibility to mix additives into the precursor solution for controlling the film formation [14, 15] as well as to implement passivation strategies for improving efficiency and stability [16]. Among the solution-based methods, blade coating is a promising, scalable technique due to its excellent material usage rate (~ 95 %) and high throughput capabilities [17]. Over the past few years, Huang's group has developed various strategies to improve the film uniformity over large areas as well as the device performance by compositional engineering [18] as well as by mixing surfactants [15], dopants [19], additives [16], and solvents [20]. Recently, they discovered a substantial void fraction in the buried perovskite/hole transporting layer interface induced by trapped dimethyl sulfoxide (DMSO) during film formation. The interfacial void fraction was reduced by partially substituting DMSO with solid-state carbohydrazide to achieve 23.6 % efficient p-i-n PSC and 19.2 % efficient mini-module with an aperture area of 50 cm$^2$ by blade coating [21]. However, to dissolve the inorganic halide precursors, solvents like 2-methoxyethanol (2-ME) or alternatively, N,N-dimethylformamide (DMF) are usually used [22, 20], which have been known for their reproductive toxicity (Category 1B, H360), bearing risk to workers and environment [23, 24, 25]. Moreover, it is challenging to form conformal films on rough surfaces, needed for monolithically integrated perovskite/silicon or perovskite/CIGS tandem solar cells [26, 27]. These drawbacks could be overcome by using vapor-based techniques [28, 29, 30]. For example, Li et al. co-evaporated methylammonium iodide (MAI) and lead iodide ($PbI_2$) to obtain PSCs with PCE of 20.28 % for 0.16 $cm^2$ small-area device and 18.13 % for 21 $cm^2$ perovskite mini-module [31]. Liu et al. sequentially evaporated $PbI_2$, formamidinium iodide (FAI) and cesium iodide (CsI) to fabricate uniform small-area devices (0.09 $cm^2$) on large-area substrate (400 $cm^2$) [32]. Although, the uniformity of vapor-based techniques is impressive, it is very challenging to implement passivation strategies or controlling the perovskite composition that is essential for high performance and superior stability [33].

In this work, we report a novel and scalable fabrication process of the perovskite film that combines the merits of the scalable solution- and vapor-based deposition methods, including facile compositional engineering to broadly tune the bandgap, and more importantly the avoidance of using toxic solvents. This hybrid PVD/blade coating process involves three steps – physical vapor deposition, blade coating and thermal annealing. We systematically vary the processing conditions of the perovskite absorber to gain insights into the perovskite formation mechanism during this process. We fabricated PSC devices with blade coated charge transporting layers ($SnO_2$ and Spiro-OMeTAD) in ambient air with green solvents. These devices achieved high open-circuit voltage ($V_{OC}$) up to 1.16 V for a perovskite with an optical bandgap of 1.56 eV and a PCE up to 18.7 %, which is the highest reported efficiency of scalable and solution-based PSCs using green solvents only.

**Hybrid PVD/blade coating fabrication process for scalable perovskite films**

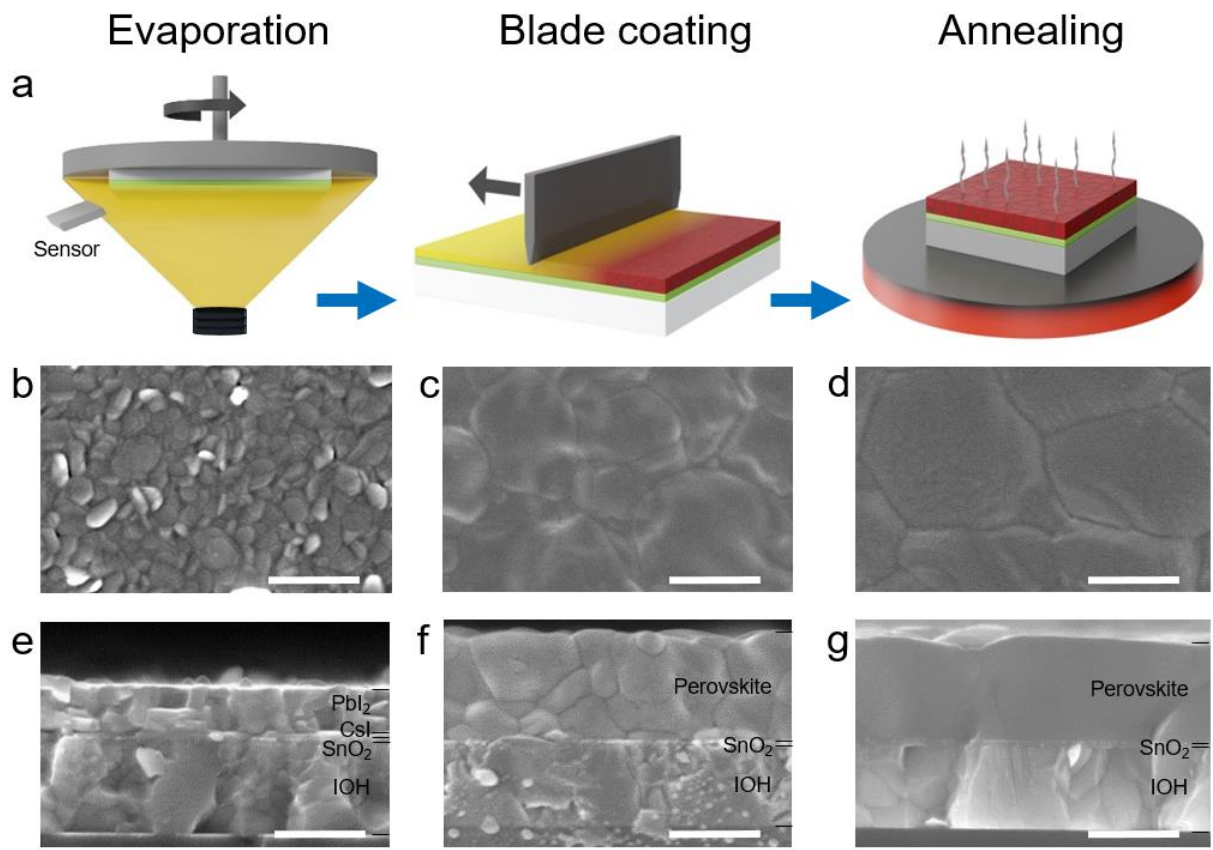

Figure 1: Hybrid PVD/blade coating process for perovskite absorber layer and morphology after each step. a) Schematic of the three-step PVD/blade coating fabrication process. SEM top and cross-section view images after each fabrication step: b, e) after evaporation, c, f) after blade coating and d, g) after thermal annealing. The scale bar is 500 nm.

**Figure 1a** shows the schematic of the scalable PVD/blade coating process to fabricate the perovskite layer. This hybrid process is composed of three steps. In the first step, the inorganic halide template is sequentially deposited by thermal evaporation. This template consists of 300 nm lead iodide ($PbI_2$) on top of 15 nm cesium iodide (CsI) layer. In the second step, the organic halide precursor solution, composed of formamidinium iodide (FAI), methylammonium bromide (MABr), and methylammonium chloride (MACl) dissolved in isopropanol, is blade coated on the inorganic halide template at substrate temperature of 65 °C. In the last step, thermal annealing at 150 °C for 15 min is performed in ambient air to obtain a compact perovskite film with large crystal grains. To study the film formation and composition, we used blade coated tin oxide ($SnO_2$) as underlying electron transporting layer and for PSC devices, we used IO:H/$SnO_2$/perovskite/Spiro-OMeTAD/Au. All deposition by blade coating are done in ambient air and with green solvents only.

**Morphology and crystallinity of the perovskite film**

In **Figure 1b-g**, scanning electron microscopy (SEM) images depict the film morphology after each step of the PVD/blade coating process. As shown in **Figure 1b**, the $PbI_2$ layer is composed of elliptical plate-shaped grains. **Figure 1e** shows a smooth and uniform inorganic halide template, which is beneficial for pinhole-free perovskite films [34]. After blade coating the organic halide solution, most of the inorganic halides are converted into α-phase perovskite even before thermal annealing, evidenced by the XRD pattern in **Figure S1**. This as-deposited perovskite film shows small perovskite grains without voids **(Figure 1c, 1f)** and its thickness is approximately twice the thickness of the inorganic halide template [35]. Subsequent thermal annealing results in large crystal grains and reduced grain boundaries **(Figure 1d)** [36] with comparable surface roughness **(Figure 1g)**. We quantified the surface roughness of the film on 5 × 5 $\mu m^2$ area with atomic force microscopy (AFM) to be 8.76 nm after evaporation, 30.2 nm after blade coating, and 34.7 nm after thermal annealing **(Figure S2)**.

**Perovskite film formation mechanism**

In this work, we focus on the processing conditions of the second step - blade coating. For a fixed gap of to 100 μm between blade and substrate, we systematically investigate the organic halide concentration, blade coating speed and substrate temperature to elucidate the influence of these parameters on the perovskite formation mechanism. First, we use organic halide solutions of different concentration ranging from 10/1/1 to 90/9/9 mg/mL of FAI/MABr/MACl, while maintaining the substrate temperature at 65 °C and the speed at 30 mm/s. The results are shown in **Figure 2** and **Figure S3**.

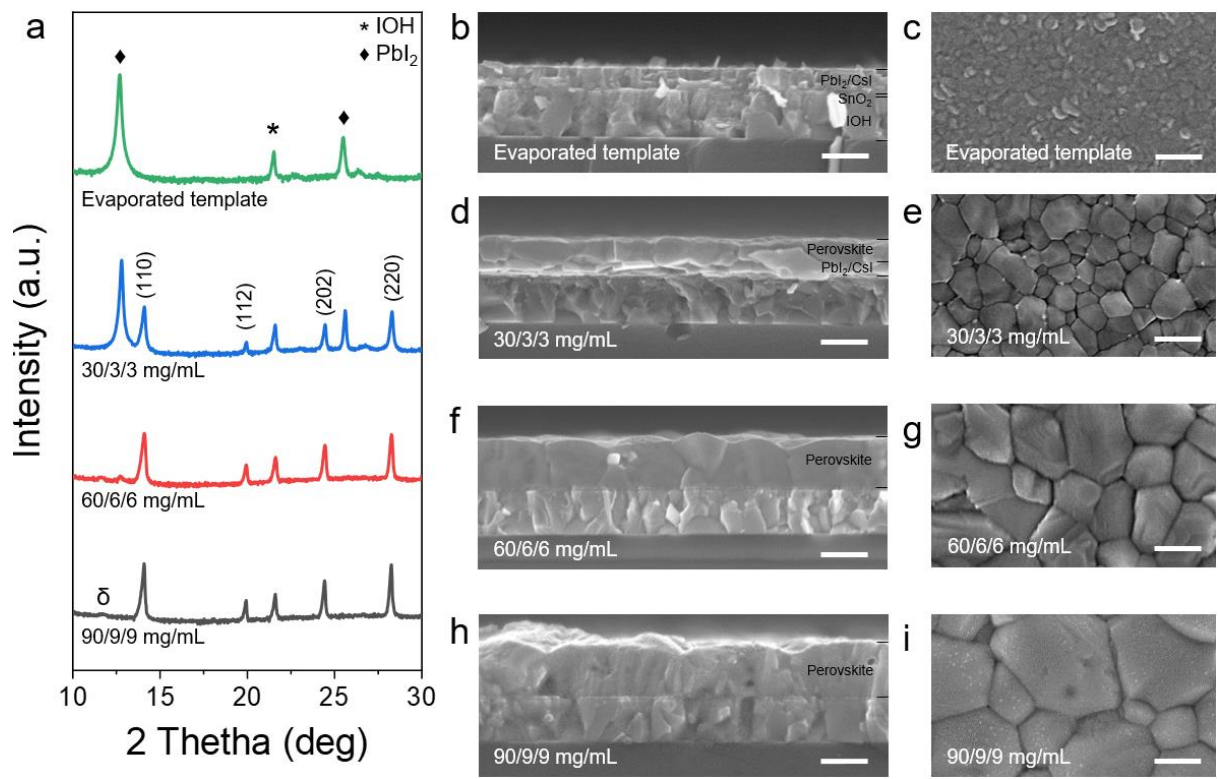

Figure 2: Evaporated inorganic halide template and perovskite films prepared by different concentrations of the organic halide solution. a) XRD patterns (log-scale) with indicated diffraction peaks of the α-phase perovskite crystal planes. $PbI_2$ is labeled by ♦ and IOH by *. b-i) SEM cross-section and top view images of b-c) inorganic halide template and perovskite films prepared by different concentrations: d-e) 30/3/3 mg/mL, f-g) 60/6/6 mg/mL and h-i) 90/9/9 mg/mL of FAI/MABr/MACl. The scale bar is 500 nm.

For low concentrations, 10/1/1 mg/mL **(Figure S3a)** and 30/3/3 mg/mL **(Figure 2a, S3a)**, the films have partially converted to perovskite. While the film obtained by 10/1/1 mg/mL appears yellowish, the film by 30/3/3 mg/mL is brownish, indicating a higher degree of perovskite conversion **(Figure S3b, S3c)**. For both concentrations, perovskite has formed in the upper region of the film, whereas residual $PbI_2$ is located in the lower region. We further detect bright dots at the grain boundaries **(Figure 2e)**, which we related to $PbI_2$. When we use concentrations between 50/5/5 mg/mL and 70/7/7 mg/mL, the film has converted to perovskite, appearing dark brown **(Figure S3d – f)**. While XRD and SEM still show a large amount of unreacted $PbI_2$ for 50/5/5 mg/mL, for 60/6/6 mg/mL, $PbI_2$ is almost fully converted to a compact perovskite layer **(Figure 2f, 2g, S3e)**. However, using 70/7/7 mg/mL, voids have formed in the perovskite **(Figure S3f)**. We suggest that during the coalescence of perovskite grains, no $PbI_2$ is trapped between adjacent perovskite grains, which could react with the organic halide precursors to prevent the formation of voids **(Figure S4)**. Further increasing the concentration to 90/9/9 mg/mL results in defective films with large crystal grains and voids as well as crystallized organic halides, visible as gray dots on the crystal grain surfaces **(Figure 2h, 2i, S3g)**. Furthermore, hexagonal perovskite δ-phase [37] has formed (2 Theta ≈ 11.8 °), making this perovskite film inappropriate for PSC devices **(Figure 2a)**.

Next, we use different speeds from 1 mm/s to 90 mm/s and blade coated the organic halide solution with concentration of 60/6/6 mg/mL FAI/MABr/MACl at 65 °C substrate temperature **(Figure S5)**. For blade coating, two coating regimes exist, the evaporation regime and the Landau-Levich regime [15]. In the evaporation regime (speed ≤ 10 mm/s), slower coating speed yields in thicker layers, i.e. a higher amount of organic halide precursors is supplied due to the increased residence time per unit length. Therefore, the diffraction peak of $PbI_2$ decreases with reduced speed (1 mm/s) and the perovskite peak intensity increases compared to 10 mm/s **(Figure S5a)**. The higher perovskite conversion degree can also be observed by the lower transparency of the sample pictures **(Figure S5b, S5c)**, showing incomplete perovskite conversion in the evaporation regime.

With coating speeds > 10 mm/s, the Landau-Levich regime is reached, where viscous forces become dominant, dragging more solution on the substrate with increasing speed [38]. Hence, the degree of perovskite conversion increases with increasing speed. Using a speed of 30 mm/s, a compact perovskite film with remnant $PbI_2$ is obtained **(Figure S5d)**. The sample shows a uniform coating. With a coating speed ≥ 50 mm/s, no remnant $PbI_2$ can be detected and the film uniformity is lost. At 50 mm/s, we observe the formation of voids in the perovskite bulk **(Figure S5e)** and the formation of δ-phase perovskite **(Figure S5a)**. Interestingly, this δ-phase perovskite has been formed immediately after blade coating at 65 °C. Further increasing the speed, results in large and coarse crystal grains with high crystallinity **(Figure S5f, S5g)**.

Lastly, we investigate the influence of different substrate temperatures on the perovskite formation **(Figure S6)**, ranging from 25 °C to 75 °. We maintain the coating speed at 30 mm/s and used an organic halide concentration of 60/6/6 mg/mL. For low substrate temperatures ≤ 40 °C, complete perovskite conversion is obtained, enabled by the slower drying rate. However, the complete conversion comes at expense of uniformity due to the coffee-ring effect, which also leads to regions of δ-phase perovskite **(Figure S6b, S6c)**. With 55 °C substrate temperature or higher, the coffee-ring effect is suppressed, resulting in uniform coatings. A substrate temperature of 65 °C shows the highest degree of perovskite while maintaining uniformity and crystallinity **(Figure S6e)**. At 75 °C, we observe decreasing perovskite conversion as the drying rate is too high, interrupting the perovskite conversion on the surface and leading to remnant $PbI_2$ on the surface (Figure S6f).

To sum up our findings, we show the perovskite film blade coated with 70 mm/s at 65 °C with an organic halide concentration of 60/6/6 mg/ in **Figure S7**. We identify three regions: δ-phase perovskite **(Figure S7d, red)**, transition **(Figure S7e, orange)** and α-phase perovskite **(Figure**

**S7f, green)** and provide the morphologies. We have shown that the organic halide precursor concentration, the coating speed and the substrate temperature play important roles in obtaining uniform perovskite films by the PVD/blade process. Next, we perform depth profiling using time-of-flight secondary ion mass spectroscopy (ToF-SIMS) and X-ray photoelectron spectroscopy (XPS) to shed light on the film composition and the cationic interdiffusion process.

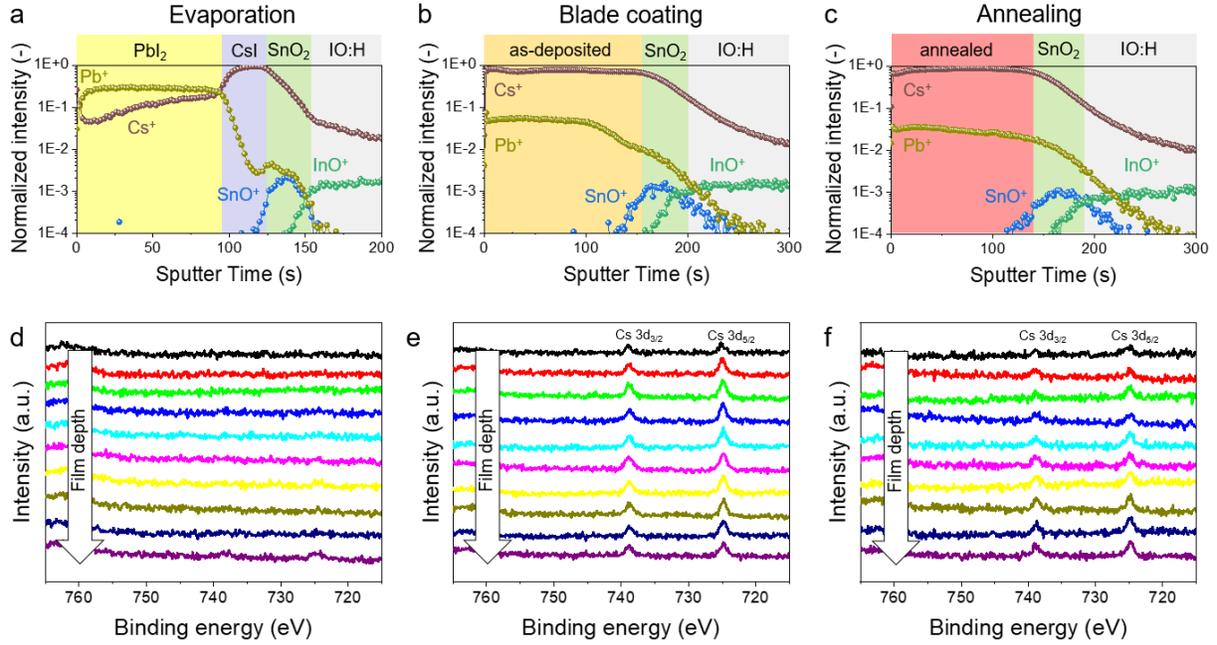

Figure 3: ToF-SIMS depth profiles of (a-c) cation species and depth profiles of XPS Cs 3d spectra at each stage of the hybrid PVD/blade coating process (d-f). The ten lines correspond to profiles at 0 nm, 10 nm, 20 nm, 30 nm, 40 nm, 80 nm, 120 nm, 160 nm, 200 nm and 300 nm depth.

**Figure 3** shows the depth profiles obtained by ToF-SIMS and XPS after each step of the PVD/blade coating process. After thermal evaporation, the bilayer of the inorganic template can be clearly differentiated **(Figure 3a)** and Cs 3d peaks in the XPS spectrum can be detected at a profile depth of 300 nm **(Figure 3d)**. After blade coating, a uniform signal of Cs cations shows that Cs species have already diffused **(Figure 3b, S8)** and confirmed by XPS spectrum **(Figure 3e)**. After thermal annealing, ToF-SIMS and XPS results are comparable as after blade coating.

**Optoelectronic properties of the perovskite film**

To evaluate the optoelectronic properties of the perovskite film by PVD/blade coating process, we use two-step spin coated perovskite film (Spin + Spin) as comparison. We show in **Note S1**, that the morphology of the evaporated inorganic halide template remains unchanged when using different amorphous substrates [39]. **Figure S9a** compares the absorbance and the photoluminescence (PL) spectra of the perovskite film for both fabrication methods. Both films have the PL peaks at 794 nm (1.56 eV). In **Figure S10**, we provide Tauc plots of these perovskite films and estimated the optical bandgap to be 1.56 eV for both absorbers [40]. We further investigate the charge recombination dynamics by time-resolved photoluminescence (TRPL) decay measurements **(Figure S9b)**. With a bi-exponential fit, we extract the fast decay lifetime $\tau_1$ and the slow decay lifetime $\tau_2$ [41]. The fitting parameters are provided in **Table S1**. The charge carrier lifetimes are comparably long for both perovskite films by PVD/blade (343 ns for $\tau_1$ and 1406 ns for $\tau_2$) and Spin + Spin (396 ns and 1148 ns), showing that perovskite films fabricated by PVD/blade achieve comparable crystal quality and photovoltaic performance **(Figure S11)**.

**Photovoltaic performance of PSCs by PVD/blade**

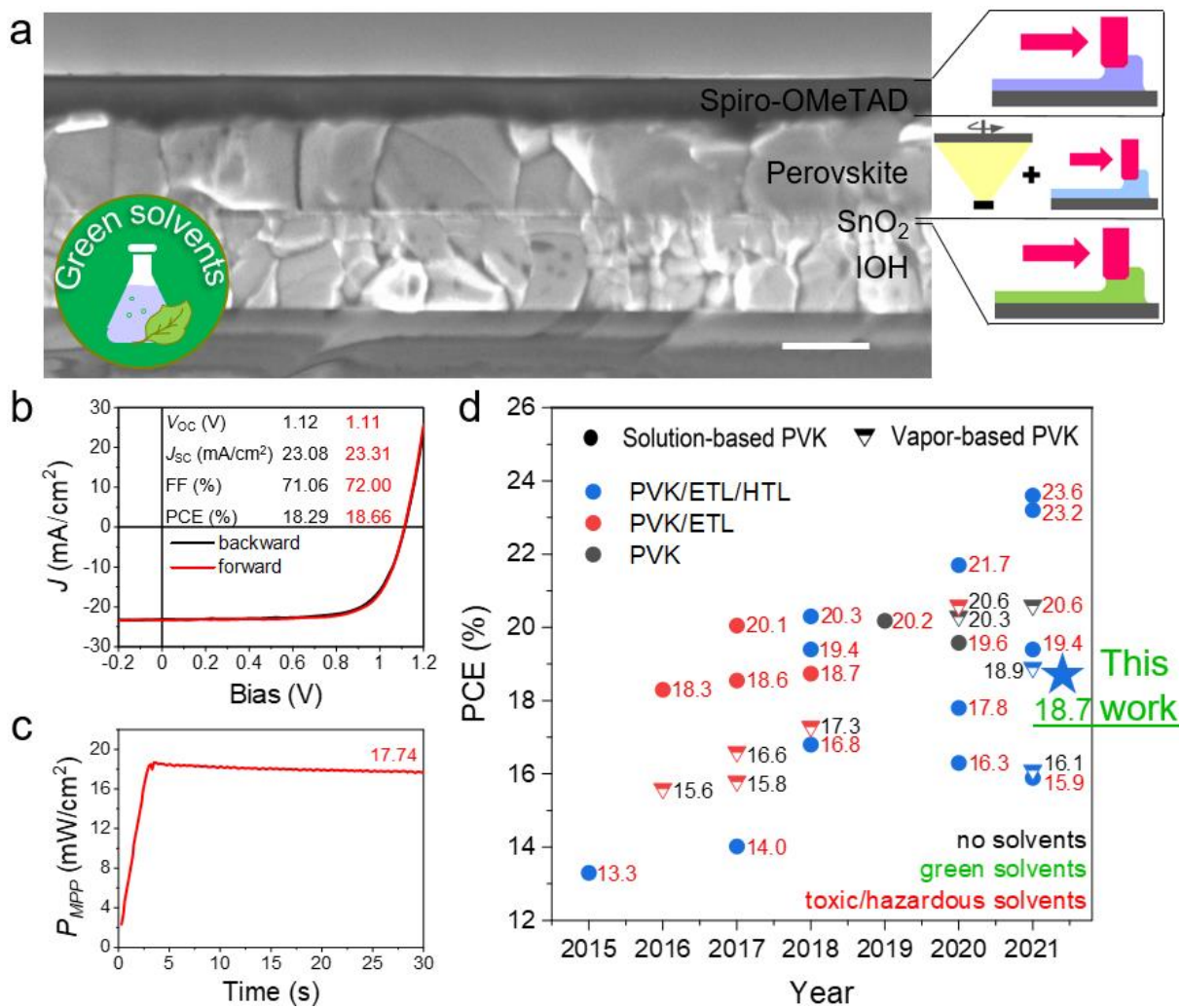

Figure 4: a) SEM cross-section view of PSC with blade coated charge transporting layers and perovskite layer by PVD/blade coating. The scale bar is 500 nm. b) Current-voltage behavior and c) MPP tracking of the champion device. d) Power conversion efficiency versus year for selected, small-area (< 0.1 cm$^2$) devices with indicated layers fabricated by scalable deposition methods (symbol color). Additionally, devices are distinguished by solution- and vapor-based perovskite absorber layer (symbol type) as well as by the use of green or toxic/hazardous in the perovskite fabrication (font color). Perovskite (PVK), electron transporting layer (ETL) and hole transporting layer (HTL).

In **Figure 4**, we evaluate the photovoltaic performance of PSCs with perovskite absorber by PVD/blade process. The n-i-p PSC devices are fabricated on 5 cm x 5 cm substrates with blade coated charge transporting layers in ambient air and green solvents, namely de-ionized water for $SnO_2$ and p-Xylene for Spiro-OMeTAD [42]. We use phenethylammonium iodide (PEAI) to passivate the perovskite layer [43]. **Figure 4a** shows a SEM cross-section image and **Figure S12** the XRD pattern of the device. The J-V measurement of the champion device reveals negligible hysteresis with an open-circuit voltage ($V_{OC}$) of 1.11 V, a short-circuit current density ($J_{SC}$) of 23.31 mA/cm$^2$ and a fill factor (FF) of 72.00 %, resulting in a PCE of 18.66 % **(Figure 4b)**. The maximum power point (MPP) tracking yields 17.74 mW/cm$^2$ after 30 s **(Figure 4c)**. This is the highest value for fully scalable PSCs with green solvents **(Figure 4d)**. Additionally, we provide photovoltaic performance statistics of 34 cells on 5 cm x 5 cm substrates to show the uniformity of the process **(Figure S13)**. On average, we obtain a $V_{OC}$ of around 1.13 V and $J_{SC}$ of 21 mA/cm$^2$. The best device showed $V_{OC}$ of 1.16 V for a bandgap of 1.56 eV, being only limited by FF and $J_{SC}$. The external quantum efficiency (EQE) measurement yields integrated $J_{SC}$ value of 21.86 mA/cm$^2$, showing further scope of improvement in the light management **(Figure S14)**.

In conclusion, we report a new and scalable hybrid perovskite fabrication process, combining thermal evaporation and blade coating using non-toxic solvents. We systematically investigate the influence of processing parameters on the perovskite formation mechanism during the PVD/blade process and elucidate the growth mechanism. We use such perovskite absorbers together with ambient air-processed, blade coated charge transporting layers to fabricate scalable PSCs with $V_{OC}$ of 1.16 V and PCE up to 18.7 % using green solvents only. With our work, we make an important contribution to reduce the performance gap between toxic and green solvent-based, scalable perovskite fabrication. PVD/blade is a promising process to fabricate high quality perovskite layers on large-area substrates. It can allow a facile implementation of passivation strategies and additives into the perovskite absorber. Additionally, it can be adapted for wide bandgap perovskite films and be eventually applied to coat textured surfaces, e.g. in tandem solar cell applications. Moreover, other meniscus-guided deposition methods can replace blade coating, such as slot-die coating or inkjet printing on flexible foils, paving the way for roll-to-roll or sheet-to-sheet manufacturing. Overall, the proposed method and its variants are easily scalable for large area manufacturing while the avoidance of toxic solvents is an additional advantage for successful commercialization of PSC technology.

ASSOCIATED CONTENT

**Supporting Information**. Additional notes on inorganic halide template and two-step fabrication methods, XRD patterns, AFM images, SEM cross-section view images, Tauc and absorption coefficient plots, table of TRPL decay lifetimes and reference list to chart.

**Notes**

The authors declare no competing financial interest.

ACKNOWLEDGMENT

This work has received funding from the European Union's Horizon 2020 research and innovation program under grant agreement No. 850937 of the PERCISTAND project, the Swiss Federal Office of Energy (SFOE, Project CIGSPSC, grant no. SI/501805-01), and the Swiss National Science Foundation (SNF, Project Bridge Power, grant no. 176552).

# Triple-cation perovskite solar cells fabricated by hybrid PVD/blade coating process using green solvents

*Severin Siegrist, Shih-Chi Yang, Evgeniia Gilshtein, Xiaoxiao Sun, Ayodhya N. Tiwari and Fan Fu\**

Laboratory for Thin Films and Photovoltaics, Empa – Swiss Federal Laboratories for Materials Science and Technology, Ueberlandstrasse 129, 8600 Duebendorf, Switzerland

*Email: fan.fu@empa.ch

EXPERIMENTAL SECTION

**Perovskite film by PVD/blade:** The inorganic halide template was sequentially deposited in a high-vacuum (< 6 x $10^{-6}$ mbar), in-house developed thermal evaporator. First, 15 nm of CsI (> 99 %, TCI) was evaporated with 0.1 nm/s, followed by 300 nm of $PbI_2$ (> 99 %, TCI) with 0.7 nm/s. Next, the organic halide precursors, formamidinium iodide (FAI, > 99.99 %, greatcellsolar), methylammonium bromide (MABr, > 99.99 %, greatcellsolar) and methylammonium chloride (MACl, > 99.0 %, Sigma-Aldrich) were mixed according to FAI/MABr/MACl 60/6/6 mg in 1ml isopropanol. The mixed solution was blade coated with 30 mm/s at 65 °C and with a gap of 100 µm in ambient air, followed by 15 min annealing at 150 °C in ambient air (~ 35 % relative humidity).

**Perovskite film by Spin + Spin:** The inorganic halide template was deposited by spin coating 1.3 M of $PbI_2$ (> 99 %, TCI) and 0.065 M of CsI (> 99 %, TCI) dissolved in mixed N,N-dimethylformamide (DMF, 99.8 %, Sigma-Aldrich) and dimethyl sulfoxide (DMSO, ≥ 99.9 %) with volume ratio of 9:1 at 1500 r.p.m. for 30 s and annealed at 70 °C for 1 min inside the glovebox. Next, the organic halide precursors, formamidinium iodide (FAI, > 99.99 %, greatcellsolar), methylammonium bromide (MABr, > 99.99 %, greatcellsolar) and methylammonium chloride (MACl, > 99.0 %, Sigma-Aldrich) were mixed according to FAI/MABr/MACl 60/6/6 mg in 1ml isopropanol. The mixed solution was spin coated with 1300 r.p.m. for 40 s in the glovebox, followed by 15 min annealing at 150 °C in ambient air (~ 35 % relative humidity).

**Solar cell fabrication:** The solar cell was fabricated with a layer structure of IOH/$SnO_2$/perovskite/PEAI/Spiro-OMeTAD/Au. First, hydrogenated indium oxide (IOH) with a sheet resistance of 10 Ω/sq were deposited on cleaned soda-lime glass substrates in a high-vacuum sputtering system (CT200, Allianceconcept) by RF sputtering of ceramic $In_2O_3$ targets (99.99 %, 10 in. diameter, SPM AG) in a mixed Ar, $O_2$, and $H_2$ atmosphere at room temperature according to Yang et al. [64]. Afterwards, the IOH film was annealed at 200 °C for 20 min in ambient air (35 - 40 % rel. humidity). Oxygen plasma treatment was performed for 5 min on the substrate before blade coating the electron transporting layer (ETL). $SnO_2$ colloid precursor (15 wt% in $H_2O$ colloidal dispersion, Alfa Aesar) was diluted in de-ionized water 1:3 by volume and blade coated with 30 mm/s at 70 °C and a gap between blade and substrate of 100 µm in ambient air. The $SnO_2$ was annealed at 150 °C for 30 min in ambient air. The perovskite with thickness around 560 nm was deposited on top of the ETL via one of the indicated two-step fabrication methods. After perovskite formation, 5 mg of phenethylammonium iodide (PEAI, 98 %, Sigma-Aldrich) per ml of isopropanol was spin coated onto the perovskite film with a spin rate of 5000 r.p.m. Next, the hole transporting layer Spiro-OMeTAD layer was blade coated with 90 mm/s at 40 °C and with a gap of 100 µm in ambient air on top of perovskite layer with a solution that contained 25 mg 2,2',7,7'-tetrakis-(N,N'-di-p-methoxyphenylamine)-9,9'-spirobifluorene (Xi'an Polymer Light Technology), 10 µl of 4-tert-butylpyridine (98 %, Sigma-Aldrich) and 6 µl of lithium bis(trifluoromethanesulfonyl)imide (Li-TFSI) solution (520 mg of Li-TFSI (99 %, Sigma-Aldrich) in 1 ml of acetonitrile (99.8 %, Sigma-Aldrich)) in 1 ml of p-Xylene ( ≥ 99 %, Sigma-Aldrich). Finally, 80 nm Au back electrode was thermally evaporated with 0.15 nm/s at a base pressure of < 6 x $10^{-6}$ mbar.

**Solar cell performance characterizations:** Photocurrent−voltage (J−V) characteristics were measured in four-contact mode at standard test conditions (100 mWcm$^{-2}$) using a Keithley 2400 source meter. A solar simulator (ABA class, LOT-QuantumDesign) was calibrated using a certified monocrystalline silicon solar cell (RS-ID-5, Fraunhofer-ISE) and was used to simulate the AM 1.5 G one sun illumination. Illumination is in superstrate configuration without external cooling of the sample. The active area is 0.1 cm$^2$ of each pixel, defined by the mask of the gold electrode. The J–V measurements were performed in both forward (from -0.2 to 1.2 V) and backward (from 1.2 to -0.2 V) directions in ambient air (~40 % relative humidity) at room temperature. No pretreatment, e.g., bias or illumination, was applied before measurement. The steady-state efficiency as a function of time was recorded using a maximum power point tracker, which adjusts the applied voltage in order to reach the maximum power point (perturb and observe algorithm). The external quantum efficiency of the devices were measured with a lock-in amplifier. The probing beam was generated by a chopped white source (900W, halogen lamp, 280 Hz) and a dual grating monochromator. The beam size was adjusted to ensure that the illumination area was fully inside the cell area. A certified single crystalline silicon solar cell was used as a reference cell. White light bias was applied during the measurement with ~ 0.1 sun intensity.

**Material characterizations**

**ToF-SIMS:** Element depth profiles were obtained with a time-of-flight secondary ion mass spectrometer (ToF.SIMS V system, ION-TOF). The primary beam was 25 keV Bi$^{3+}$ with a total current of 0.22 pA and a raster size of 50 × 50 μm$^2$. For detection of negative ions, Cs$^+$ ions were used with 500 eV ion energy, 25 nA pulse current on a 300 × 300 μm$^2$ raster size to bombard and etch the film. For detection of positive ions, sputtering was performed using O$_2^-$ ions at 2 keV ion energy, 180 nA pulse current over a raster size of 300 × 300 μm$^2$. The data were plotted with the intensity for each signal normalized to the total counts of the signal.

**XPS.** X-ray photoelectron spectroscopy was performed using a Quantum 2000 system from Physical Electronics with a monochromatic Al K$_\alpha$ source (1486.6 eV, spot size, 100 μm) and a base pressure below 8 x 10$^{-8}$ mbar. No surface cleaning was performed. Detailed high-resolution scans of I 3d, S 2p, N 1s, Cl 2p, O 1s, Br 3d, C 1s, Cs 3d, Pb 4f were recorded with an energy step size of 0.125 eV and a pass energy of 29.35 eV during depth profile sputtering by $Ar^+$ sputtering. The first five cycles (with the first measurement performed directly on the surface) were performed with 0.5 kV energy after 30 s sputtering each, followed by five more cycles: three with 2 min intervals and 2 more with 4 min intervals with the same energy. The sputtering rate was estimated to be around 15-20 nm/min which corresponds to 250-300 nm total depth of the performed analysis. With the use of MultiPak processing software all data were background subtracted.

**TRPL/PL:** Time-resolved photoluminescence (TRPL) and steady-state photoluminescence (PL) were measured using a FluoTime 300 unit coupled into a MicroTime 100 system from PicoQuant. For TRPL measurement, a 639 nm pulsed laser with < 100 ps pulse width was used as an excitation source. The spot size was measured using a NanoScan2 beam profiler resulting in ~ 130 mm diameter. The excitation was $3\times10^{11}$ photons $cm^{-2}$ $pulse^{-1}$ and the pulse frequency was 0.1 MHz. For the fitting procedure, a bi-exponential decay was used and the first 20 ns after the signal peak were neglected.

**AFM:** The AFM characterization was carried out using an AFM microscope (Bruker ICON3) in the air. A silicon nitride tip (ScanAsyst-air) with a radius of 10 nm was used as the probe. The cantilevers' spring constant and resonant frequency are 0.4 N/m and 70 kHz, respectively.

**UV-Vis:** Reflectance and transmittance measurements were performed using a Shimadzu UV-Vis 3600 spectrophotometer equipped with an integrating sphere. The reflectance data were corrected for the instrumental response stemming from diffuse and specular reflections both on the sample, and the reflectance measurements were carried out in a wavelength range from 300 to 1500 nm.

**XRD:** X-Ray diffraction patterns were measured on an X'Pert Pro in Bragg–Brentano geometry using Cu $K_{\alpha 1}$ radiation ($\lambda = 1.5406$ Å), scanning from 5 to 60 ° (2θ) with a step interval of 0.0167°.

**SEM:** The SEM micrographs were taken with a Hitachi S-4800 Scanning Electron Microscope using 5–10 kV acceleration voltage. A thin layer (1 nm) of Pt was coated on top of the samples to avoid charging effects.

NOTES

**Note S1: Morphology of inorganic halide template with different amorphous substrates**

Fu et al. have shown that the crystal characteristics of the substrate influence the morphology of the evaporated inorganic template [1]. While amorphous substrates yield in compact $PbI_2$ layers, poly-crystalline substrates lead to porous, plate-like $PbI_2$ layers. We use this finding to deposit perovskite films on amorphous substrates to avoid further material restrictions for the perovskite fabrication method by PVD/blade. We provide SEM cross-section images of the inorganic halide template on different amorphous materials, including a) soda-lime glass (SLG), b) IOH and c) colloidal $SnO_2$ nanocrystals to support the versatility of this method. As depicted in **Figure S15c-d**, the morphology of the evaporated inorganic halide template is similar for each of these substrates and in all cases, we obtain a compact inorganic template. Additionally, the XRD spectra of these films show a comparable crystallinity of the evaporated inorganic template with a preferred orientation of (001) plane over (101) **(Figure S15a)**. Therefore, we assume a similar infiltration mechanism for the organic halide precursor solution into the inorganic halide template. Hence, we expect a comparable perovskite film formation by PVD/blade for any amorphous substrate, including inert quartz.

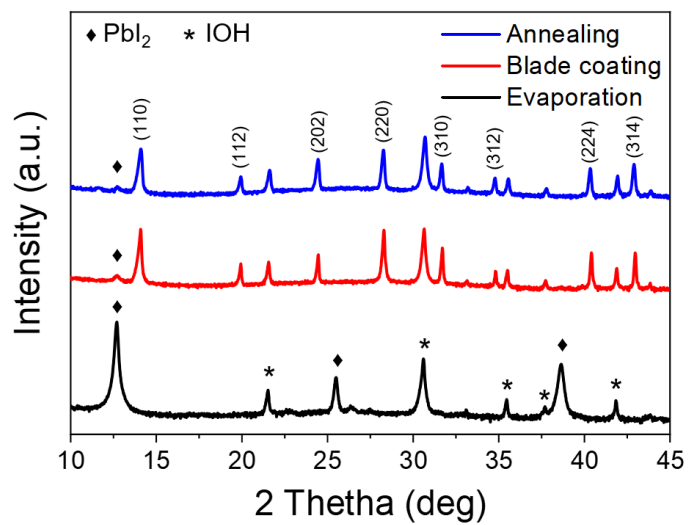

Figure S1: XRD patterns (log-scale) of the film after thermal evaporation, blade coating and annealing with indicated perovskite crystal plane orientations. $PbI_2$ is labeled by ♦ and IOH by *.

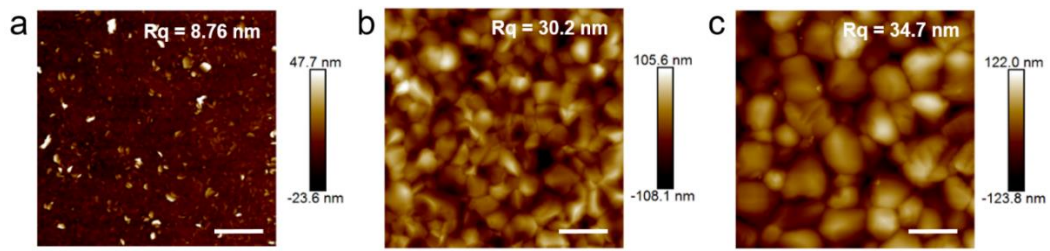

Figure S2: AFM height images of the film surface after a) evaporation, b) blade coating and c) annealing. The scale bar is 1 μm.

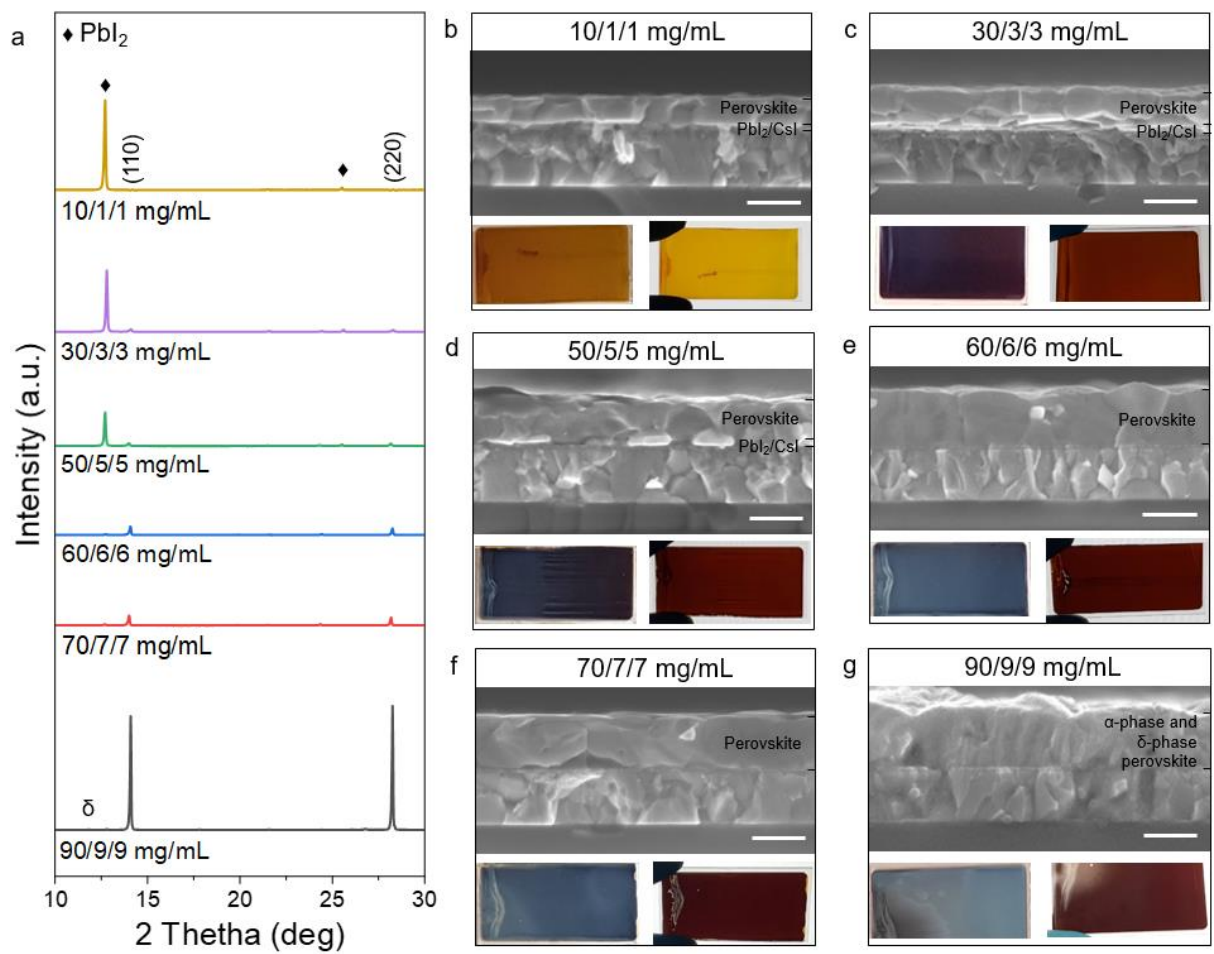

Figure S3: Perovskite films fabricated with different concentrations of the organic halide precursors (FAI/MABr/MACl in mg/mL) in the solution. a) XRD pattern with indicated diffraction peaks of the α-phase perovskite crystal planes. The δ-phase perovskite is labeled by δ and PbI$_2$ by ♦. b-g) SEM cross-section view images for each concentration with sample pictures taken in foreground and in background illumination mode. The scale bar is 500 nm.

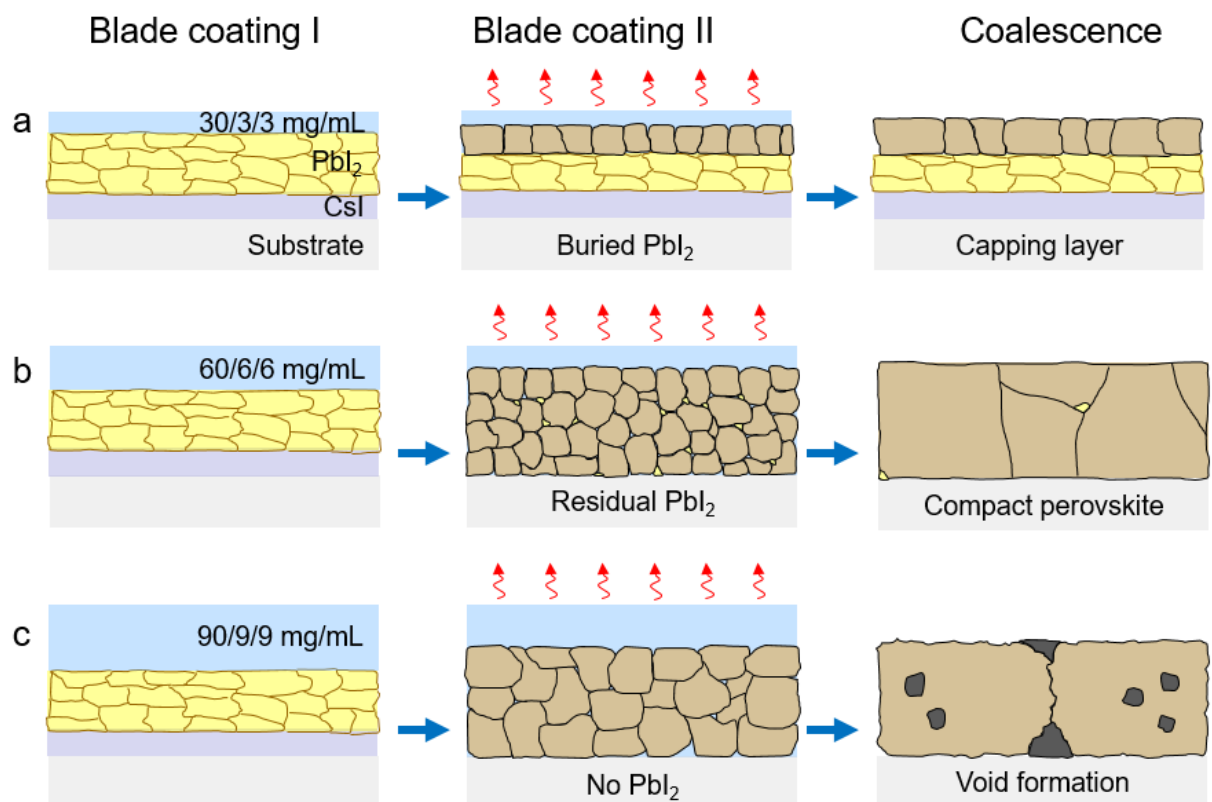

Figure S4: Influence of organic halide precursor concentration on the degree of perovskite conversion of the film. a) Formation of perovskite capping layer when blade coating of 30/3/3 mg/mL FAI/MABr/MACl on the inorganic halide template. b) Almost fully converted perovskite film with unreacted PbI$_2$ residues for 60/6/6 mg/mL. c) Formation of defective, but highly crystalline perovskite film with voids in the bulk and rough crystal grains, when using 90/9/9 mg/mL of organic halide precursors. The amount of organic halide precursors is represented by the solution thickness (blue).

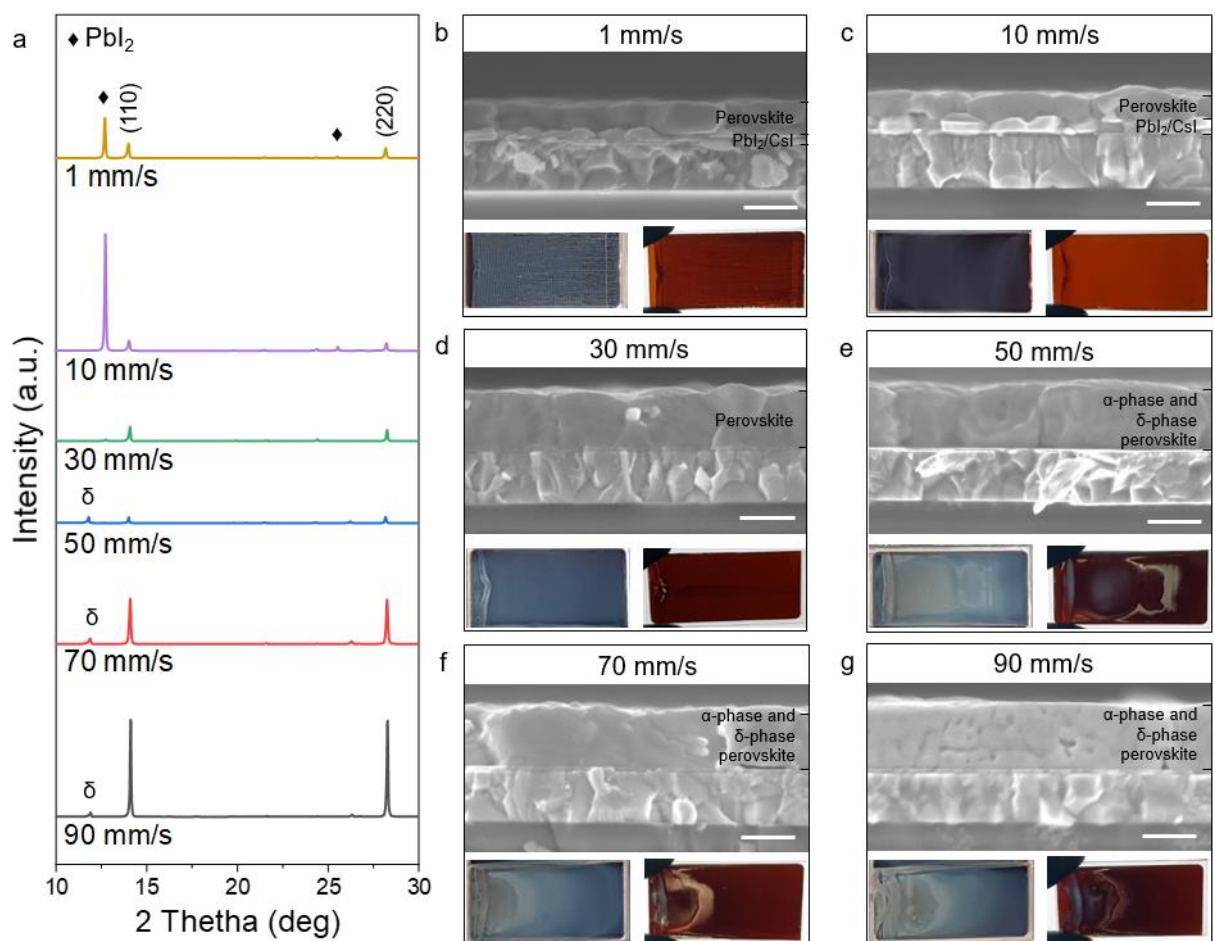

Figure S5: Perovskite films fabricated with different blading speeds. a) XRD pattern with indicated diffraction peaks of the α-phase perovskite crystal planes. The δ-phase perovskite is labeled by δ and PbI$_2$ by ♦. b-g) SEM cross-section view images for each speed with sample pictures taken in foreground and in background illumination mode. The scale bar is 500 nm.

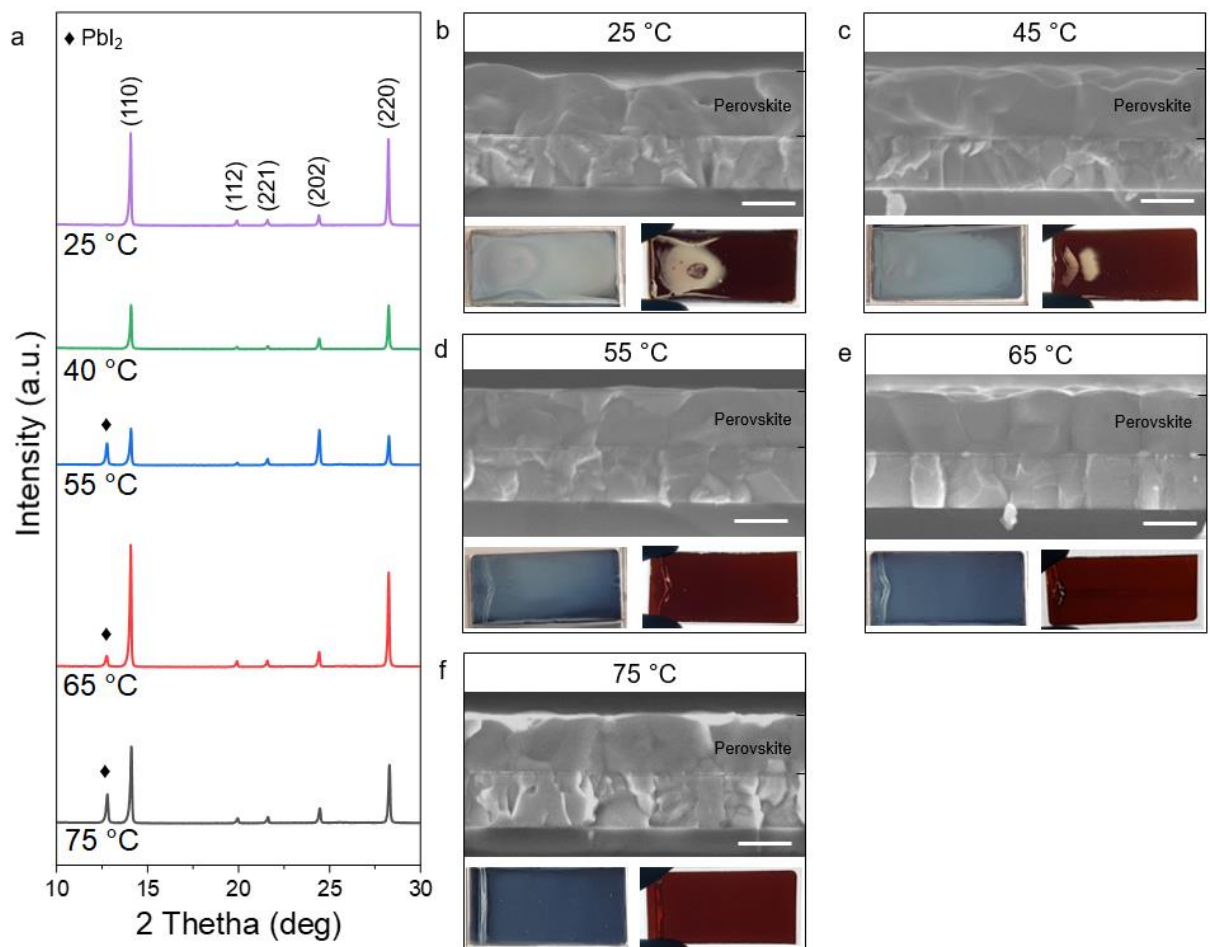

Figure S6: Perovskite films fabricated at different substrate temperatures. a) XRD pattern with indicated diffraction peaks of the α-phase perovskite crystal planes. PbI$_2$ is labeled by ♦. b-f) SEM cross-section view images for each substrate temperature with sample pictures taken in foreground and in background illumination mode. The scale bar is 500 nm.

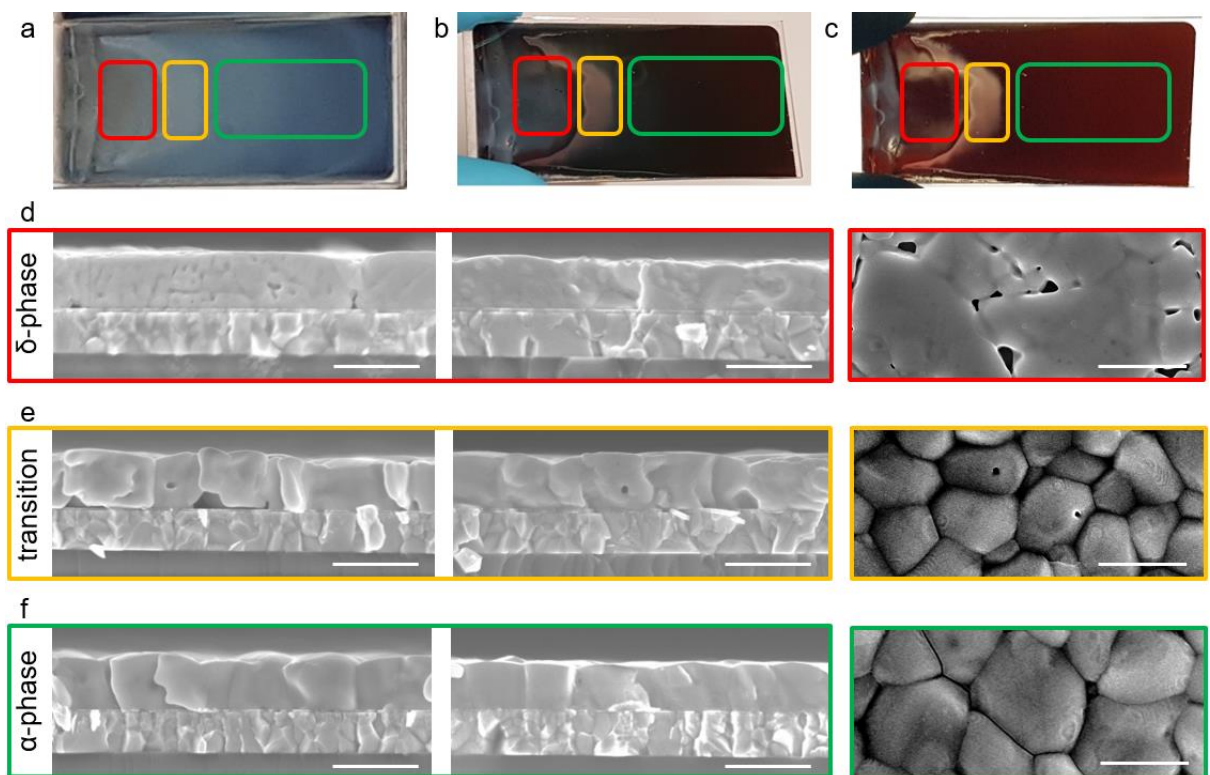

Figure S7: Perovskite crystal phases and morphology that occur, when the processing parameters are not matching, i.e. the balance of inorganic halide precursors and organic halide precursors is not given. Sample pictures of the perovskite film blade coated with 70 mm/s at 65 °C with an organic halide concentration of 60/6/6 mg/mL of FAI/MABr/MACl in foreground illumination mode: a) front side and b) backside of the sample as well as in c) background illumination mode. SEM cross-section and top view images of the d) δ-phase perovskite region, e) transition region and f) α-phase perovskite region. The scale bar is 1 µm.

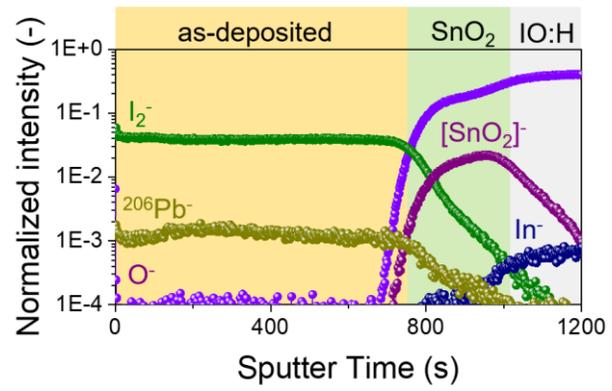

Figure S8: ToF-SIMS depth profile of negative species of the as-deposited film.

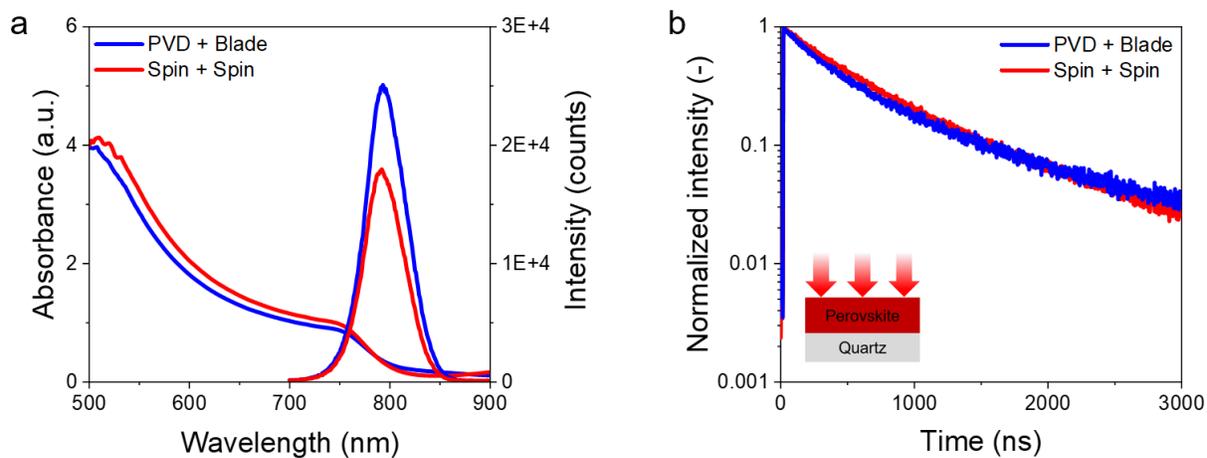

Figure S9: Optoelectronic properties of perovskite films fabricated by the PVD/blade process (PVD + Blade) and by two-step spin coating (Spin + Spin) on quartz. a) Ultraviolet-visible (UV-Vis) absorbance and photoluminescence (PL) spectra, b) time-resolved photoluminescence (TRPL) decays. The decay lifetimes are obtained by fitting the measurement with the bi-exponential decay function $I(t) = A_1 \exp\left(-\frac{t}{\tau_1}\right) + A_2 \exp\left(-\frac{t}{\tau_2}\right)$. The fitting parameters are provided in Table S1.

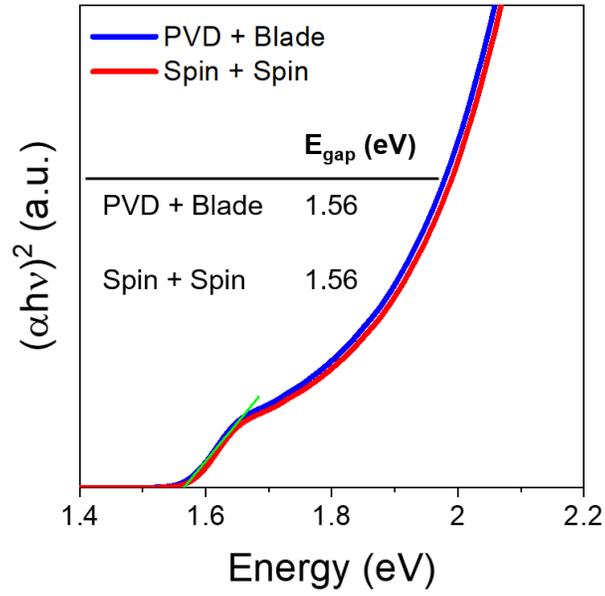

Figure S10: Tauc plot as a function of $(\alpha h\nu)^2$ versus energy of the perovskite films by the PVD/blade process (PVD + Blade) and by two-step spin coating (Spin + Spin) on quartz. The extracted bandgap energy is 1.56 eV for both methods.

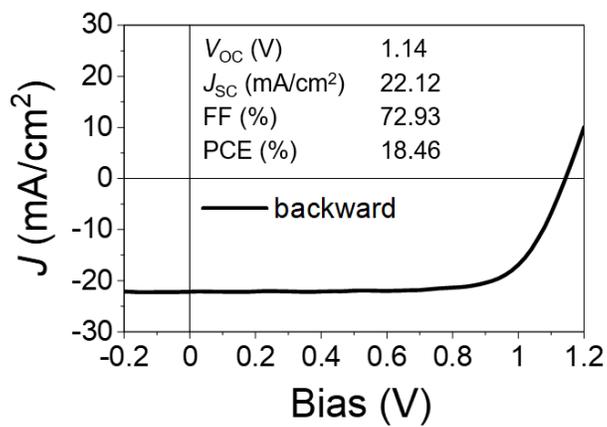

Figure S11: Photovoltaic performance of PSC with two-step spin coated perovskite absorber layer (Spin + Spin) and blade coated charge transporting layers.

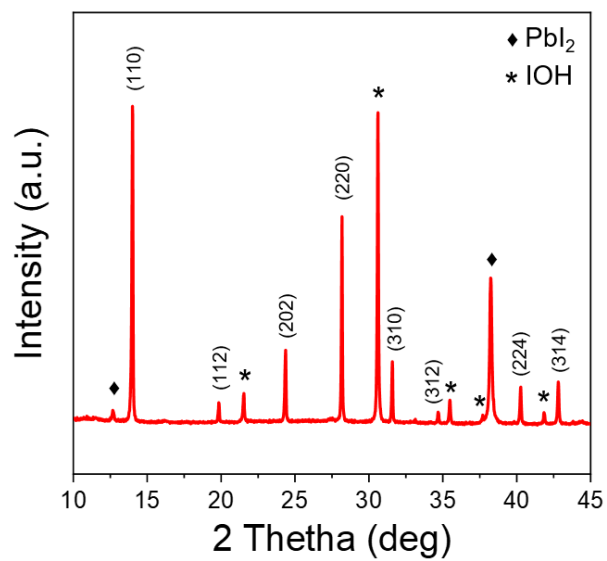

Figure S12: XRD pattern of PSC device. Perovskite crystal planes are indexed. PbI$_2$ is labeled by ♦ and IOH by *.

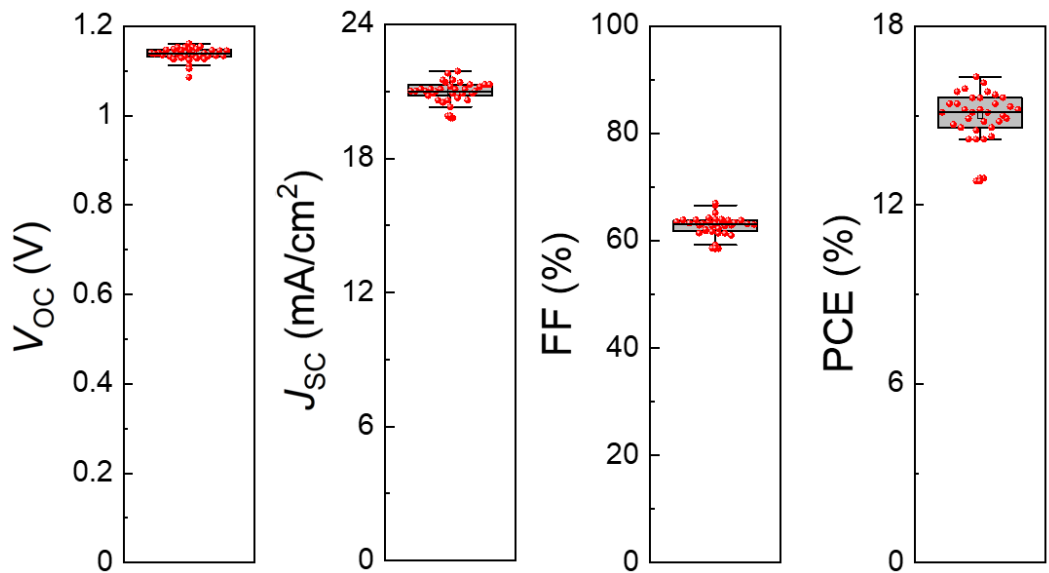

Figure S13: Photovoltaic performance overview of 36 PSC devices on 5 cm x 5 cm sample with blade coated charge transporting layers and perovskite fabricated by PVD/blade process.

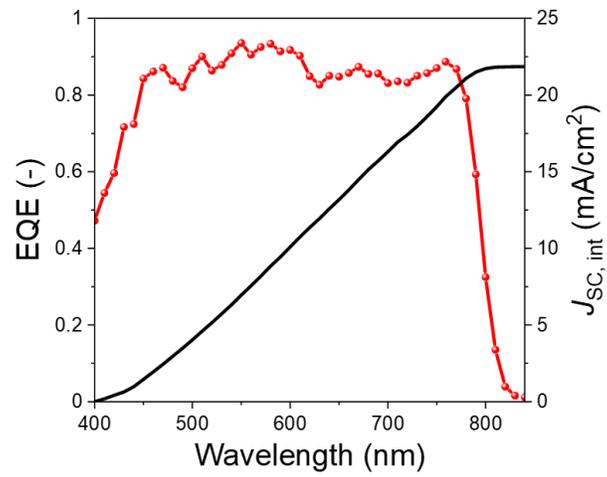

Figure S14: EQE spectrum with integrated $J_{SC}$ value of 21.86 mA/cm$^2$ for the PSC device.

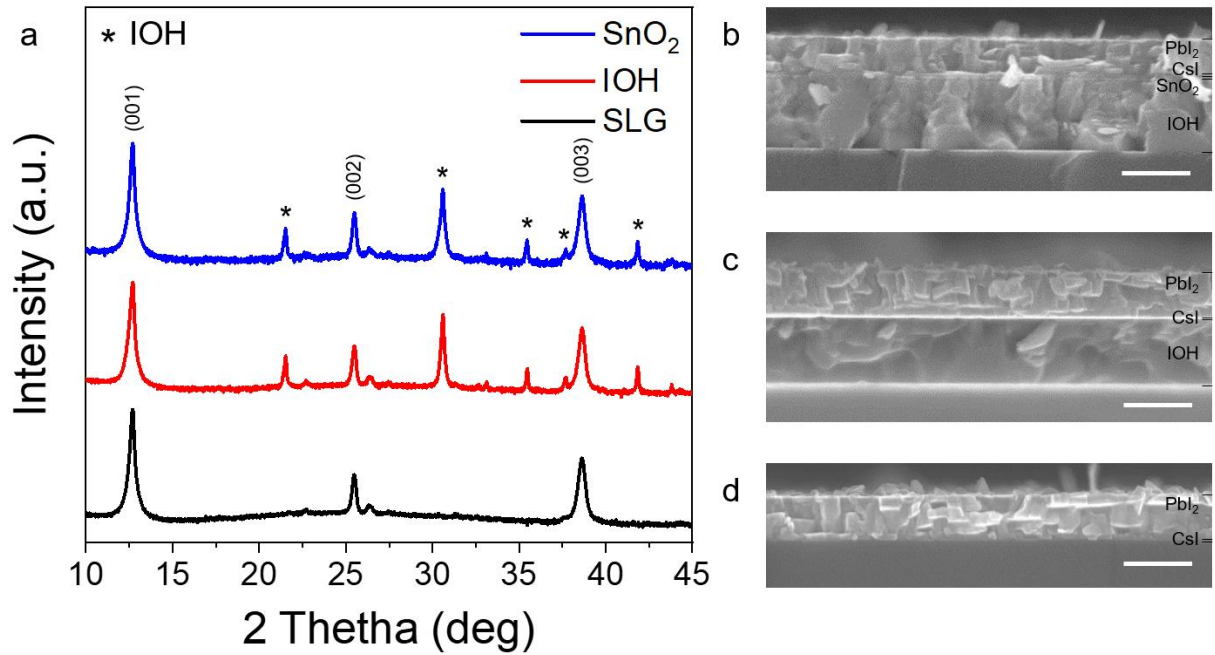

Figure S15: XRD patterns of the inorganic halide template ($PbI_2$/CsI), sequentially evaporated on $SnO_2$ (blue), on IOH (red) and on soda-lime glass (SLG, black). a) $PbI_2$ crystal planes are indexed and IOH is labeled by *. Corresponding SEM cross-section view images of the inorganic halide template on b) $SnO_2$, c) IOH and d) SLG. The scale bar is 500 nm.

Table S1: Fitting parameters of the bi-exponential decay function of the time-resolved photoluminescence (TRPL) measurement of the perovskite films obtained by PVD/blade process and by two-step spin coating. Average decay time $\tau$ was calculated according to: $\tau = \frac{A_1\tau_1 + A_2\tau_2}{A_1 + A_2}$ [6].

| Method | $A_1$ | $\tau_1$ (ns) | $A_2$ | $\tau_2$ (ns) | $\tau$ (ns) |
|---|---|---|---|---|---|
| PVD + Blade | 2094.29 | 342.52 | 737.74 | 1405.7 | 619.48 |
| Spin + Spin | 3107.67 | 395.59 | 1603.44 | 1148.4 | 651.81 |

Table S2: Summary of the chart in Figure 4d with selected work of perovskite solar cells with indicated layers fabricated by scalable deposition methods.

| YEAR | SCALABLE LAYERS | PCE | STACK | REF |
|---|---|---|---|---|
| 2019 | PVK | 20.2 | FTO/SnO2/PVK/Spiro-OMeTAD/Ag | 1 |
| 2020 | PVK | 19.6 | FTO/SnO2/PVK/Spiro-OMeTAD/Au | 2 |
| 2020 | PVK | 20.3 | ITO/SnO2/PCBM/PVK/Spiro-OMeTAD/Au | 3 |
| 2021 | PVK | 20.6 | ITO/MeO-2PACz/PVK/PCBM/BCP/Ag | 4 |
| 2016 | PVK/ETL | 18.3 | FTO/c-TiO2/PVKPTAA/Au | 5 |
| 2016 | PVK/ETL | 15.6 | FTO/c-TiO2/m-TiO2/PVK/Spiro-OMeTAD/Au | 6 |
| 2017 | PVK/ETL | 20.1 | FTO/TiO2/PVK/PTAA/Ag | 7 |
| 2017 | PVK/ETL | 16.6 | FTO/c-TiO2/PVK/Spiro-OMeTAD/Au | 8 |
| 2017 | PVK/ETL | 18.6 | FTO/c-TiO2/m-TiO2/PVK/Spiro-OMeTAD/Au | 1 |
| 2017 | PVK/ETL | 15.8 | FTO/C60/PVK/Spiro-OMeTAD/Ag | 9 |
| 2018 | PVK/ETL | 17.3 | FTO/SnO2/PVK/Spiro-OMeTAD/Au | 10 |
| 2018 | PVK/ETL | 18.7 | ITO/TiO2/PVK/Spiro-OMeTAD/Au | 11 |
| 2020 | PVK/ETL | 20.6 | ITO/MeO-2PACz/PVK/C60/BCP/Cu | 12 |
| 2015 | PVK/ETL/HTL | 13.3 | FTO/c-TiO2/m-TiO2/PVK/P3HT/Au | 13 |
| 2017 | PVK/ETL/HTL | 14.0 | FTO/c-TiO2/m-TiO2/m-ZrO2/PVK/Carbon | 14 |
| 2018 | PVK/ETL/HTL | 20.3 | ITO/PTAA/PVK/C60/BCP/Cu | 15 |
| 2018 | PVK/ETL/HTL | 19.4 | FTO/c-TiO2/PVK/Spiro-OMeTAD/Au | 16 |
| 2018 | PVK/ETL/HTL | 16.8 | FTO/c-TiO2/PVK/Spiro-OMeTAD/Au | 17 |
| 2020 | PVK/ETL/HTL | 17.8 | ITO/SnO2/SAM/PVK/PDCBT/Ta-WOx/Ag | 18 |
| 2020 | PVK/ETL/HTL | 21.7 | FTO/TiO2/PVK/PCBM/Ag | 19 |
| 2020 | PVK/ETL/HTL | 16.3 | ITO/SnO2/PVK/Spiro-OMeTAD/Au | 20 |
| 2021 | PVK/ETL/HTL | 23.6 | ITO/PTAA/PVK/C60/BCP/Cu | 21 |

| 2021 | PVK/ETL/HTL | 18.9 | FTO/TiO2/PVK/MoO3/NPB/Au | 22 |
| 2021 | PVK/ETL/HTL | 23.2 | ITO/PTAA/PVK/C60/BCP/Cu | 23 |
| 2021 | PVK/ETL/HTL | 18.7 | IOH/SnO2/PVK/Spiro-OMeTAD/Au | This work |
| 2021 | PVK/ETL/HTL | 15.9 | FTO/c-TiO2/mp-TiO2/PVK/Spiro-OMeTAD/Ag | 24 |
| 2021 | PVK/ETL/HTL | 19.4 | ITO/SnO2/PVK/HTL/Ta-WOx/Au | 25 |
| 2021 | PVK/ETL/HTL | 16.1 | TCO/CuPc/PVK/C60/BCP/Ag | 26 |

REFERENCES of Table S2

Surfactant-Assisted Room-Temperature Coating of Efficient Perovskite Solar Cells. *Joule* **2020**, *4* (11), 2404–2425. https://doi.org/10.1016/j.joule.2020.09.011.